\documentclass{article}
\usepackage{color}

\def\t{\theta}

\def\vf{\varphi}

\def\be{\begin{equation}}
\def\ee{\end{equation}}
\def\arr{\begin{array}{rll}}
\def\ea{\end{array}}
\def\bea{\begin{eqnarray}}
\def\eea{\end{eqnarray}}

\def\N2{$N{=}2$}

\def\>{\rangle}
\def\<{\langle}
\def\+{\dagger}
\def\={\ =\ }

\begin{document}
\renewcommand{\thefootnote}{\fnsymbol{footnote}}
\setcounter{page}{1}
\begin{center}
{\Large\bf Near-horizon dynamics of particle
 in  extreme Reissner-Nordstr\"om and Cl\'ement-Gal'tsov  black hole backgrounds: action-angle variables}\\

\vspace{0.5 cm} {\large
Armen Saghatelian}
\end{center}
{\sl Yerevan State University,
1 Alex Manoogian St., Yerevan, 0025, Armenia}

\begin{abstract} \noindent
We analyze the periodic motion in the conformal mechanics describing the particles moving near
the horizon of extreme Reissner-Nordstr\"om
and axion-dilaton (Cl\'ement-Gal'tsov) black holes.
For this purpose we extract the (two-dimensional) compact (``angular") parts of these systems and construct their action-angle variables.
In the first case we get the well-known spherical Landau problem, which possesses hidden $so(3)$ symmetry,
 while in the latter case the system does not have hidden constant of motion. In both cases we indicate
 the existence of ``critical points", separating the regions of periodic motions with qualitatively different properties.
\end{abstract}

\vspace{0.5cm}

%

\renewcommand{\thefootnote}{\arabic{footnote}}
\setcounter{footnote}0

\section{Introduction and summary}~

The black hole solutions allowed in supersymmetric field
theories have an extremality property, that is, the inner and outer horizons
 of the black hole coalesce. In this case one can pass to the near-horizon limit, which brings us to new solutions of Einstein equations.
 In this limit (near-horizon extreme black hole) the solutions become conformal invariant.
The conformal invariance was one of the main reasons why the extreme black holes have
been payed so much attention to for the last fifteen years. Indeed, due to conformal
invariance black hole solutions are
a good research area for studying conformal field theories and
AdS/CFT correspondence (for the recent review see \cite{bkls}). The simplest way to research this type of
configurations is to study the motion of a (super)particle in this
background. The first paper that considered such a problem is
\cite{cdkktp}, where the motion of particle near horizon of extreme
Reissner-Nordstr\"om black hole has been considered. Later similar
problems in various extreme black hole backgrounds were studied by
several authors (see \cite{ikn,galaj} and refs therein). It is
obvious, that particle moving on conformal-invariant background
 inherits the
property of (dynamical) conformal symmetry, that is, one can present
additional generators $K$ and $D$, which form, together with the
Hamiltonian $H$ , the conformal algebra $so(2,1)$ : \be
\label{confalg} \{H,D\}=H, \quad \{H,K \}=2D, \quad \{D,K \} =K. \ee

On the other hand from general reasoning (in the spirit of Darboux's
theorem) one can state that a conformal mechanics can be presented
in a non-relativistic ``canonical" form \be \label{dec}
H=\frac{p^2_R}{2}+\frac{2 {\cal  I }(u)}{R^2},\qquad
\Omega=dp_R\wedge dR+\frac 12\omega_{\alpha
\beta}(u)~du^\alpha\wedge du^\beta. \ee where $R$ and $p_R$ are the
effective radial coordinate and the
 momentum,  and ${\cal I}=HK-D^2$ is the Casimir element of $so(2,1)$ algebra.
 This means that all the characteristics of conformal mechanics are encoded in such (angular)
  system ($\frac 12 \omega_{\alpha\beta}du^\alpha\wedge du^\beta, {\cal I}(u)$) (see e.g.\cite{cuboct}, \cite{hkln},\cite{hlns}).
For instance, the angular
 part of the rational $A_N$--Calogero
model leads to the superintegrable multi--center Higgs oscillator on
an $(N-1)$--dimensional sphere \cite{cuboct}.  One can also reverse
the argument and construct new integrable conformal mechanical
systems starting from known ones\cite{hlnsy}. Conformal basis also
provides a useful tool for supersymmetrization of conformal
mechanics. For instance, given ${\cal N}=4$ supersymmetric extension of the
Hamiltonian system $(\frac 12 \omega_{\alpha\beta}du^\alpha\wedge
du^\beta, {\cal I}(u))$ allows to construct immediately superconformal
mechanics with $D(1,2;\alpha)$ symmetry \cite{hkln}.

 However,  there is no general canonical transformation
known which transforms arbitrary conformal  mechanics to the form
(\ref{dec}). For the particular case of the near-horizon motion of
the particle in the extreme Reissner-Nordstr\"om background such transformation has been suggested
 in \cite{bgik}, while recently it was extended to the case of general four-dimensional
near horizon extreme black hole in \cite{gn}.
 In the latter the authors, taking  into account the integrability of
 this system, suggested the generic canonical transformation,
 assuming  that the angular system ($\frac 12 \omega_{\alpha\beta}du^\alpha\wedge du^\beta, {\cal I}(u)$) is
 formulated in action-angle variables.
  They exemplified  their scheme,  constructing the action-angle variables for a neutral particle
  near the horizon of extreme
  Reissner-Nordstr\"om black hole, as well as discussed the case of the charged particle near the horizon
  of extreme  axion-dilaton (Cl\'ement-Gal'tsov)  black hole\cite{cg},
  without actually constructing the action-angle variables for the second
  system. Then, the action-angle
formulation for the angular part of a near-horizon  particle
dynamics in the extreme Kerr black hole background has also been
presented \cite{bny}.

The purpose of current paper is to construct  the action-angle
variables for the angular parts of the following two exactly
solvable systems:
\begin{itemize}
\item
Charged particle moving near the horizon of extreme Reissner-Nordstr\"om  black hole  with magnetic charge (unlike the neutral particle in \cite{gn})
\item
Particle moving near the horizon of extreme Cl\'ement-Gal'tsov black hole (explicit expressions)
\end{itemize}

The study of this problems not only fills the gap in the paper \cite{gn},
 but also presents its own interest.
 Namely,
 \begin{itemize}
 \item The construction of action-angle variables is the most sequential method
 of the study of periodic motion. Particularly,  formulation of the system in action-angle variables
 explicitly indicates the existence of hidden symmetries in the system.

  \item Action-angle variables give a precise indication of the (non)equivalence of
  different integrable systems
   \item
  Finally,  action-angle variables form a bridge between classical and quantum-mechanical systems, and allows
  to perform semiclassical quantization of the system.
\end{itemize}
  Unfortunately, this method has not been paid enough attention by the theoretical physics community during the last several decades.
Recently, in a series of papers  an attempt was made to show
the effectiveness of action-angle
 variables in problems of modern theoretical physics \cite{hlnsy,bny,bnsy,aa}.

Let us reinforce the above
mentioned observations on the near-horizon dynamics of particle in the
background of extreme Kerr black hole \cite{bny}: the use of
action-angle variables allowed to find there a critical point
$|p_\varphi|=2 mM  $ (with $m$ being the mass of the probe particle,
$ M$ being mass of extreme Kerr black hole), where the trajectories become
closed. We will show that there are similar singular points in the
dynamics of charged particle moving near the horizons of extreme
Reissner-Nordstr\"om and Cl\'ement-Gal'tsov black holes. They are defined
by the relation $p_\varphi=\pm s$, where  $s=ep$ for the Reissner-Nordstr\"om case (with $e$  being an electric charge of
probe particle, and $p$ being  magnetic charge of extreme Reissner-Nordstr\"om black hole), and $s=e$ for the case of extreme
Cl\'ement-Gal'tsov black hole (with $e$ being an effective electric charge of probe particle). \\[18pt]
The paper is organized as follows:

In the Second Section we describe, following Ref. \cite{gn}, the
canonical transformation that leads  the Hamiltonian of particle
moving near the horizon  of four-dimensional  extreme black hole,  to
the canonical form (\ref{dec}). Then we present the procedure of
constructing the action-angle variables for this
 particular case.

In the Third Section we construct the action-angle variables for the
angular  part of a charged particle moving near the horizon of Reissner-Nordstr\"om
extreme black hole. We show that the angular part is equivalent to the spherical
Landau problem, and has a hidden constant of motion. We find a ``critical point" that divides the different phases
of effective periodic  motion.

In the Fourth Section we discuss a charged particle moving near the
horizon of extreme Cl\'ement-Gal'tsov
  black hole.  We construct the action-angle variables for the angular  part
  of this system and use them to analyze the properties of periodic motion. In contrast with Reissner-Nordstr\"om
  case, the system does not possess hidden constant of motion.
  We find a critical  point that divide the phases (both effectively two-dimensional ones) of rotations  in opposite directions.

\section{Canonical transformation \cite{gn}}

Conformal mechanics associated with the near horizon geometry of an
extremal black hole in four dimensions is described by the triple
\be \label{ham} H=r \left( \sqrt{{(r p_r)}^2 + L(\t,p_\t,p_\varphi)}
-q(p_\varphi) \right), \quad K=\frac{1}{r} \left( \sqrt{{(r p_r)}^2
+  L(\t,p_\t,p_\varphi)}+q(p_\varphi) \right), \quad D=r p_r, \ee
which involves the Hamiltonian $H$, the generator of dilatations
$D$, and the generator of special conformal transformations $K$.
Under the Poisson bracket they form conformal algebra $so(2,1)$
(\ref{confalg}).

The functions $L(\t,p_\t,p_\varphi)$ and $q(p_\varphi)$ entering Eq. (\ref{ham}) depend
on the details of a particular black hole under consideration: see
\cite{bgik,galaj} for the near horizon extremal Reissner-Nordstr\"om
black hole, \cite{cg1} for the rotating extremal dilaton--axion black
hole,  \cite{g2} for the extremal Kerr  solution, and \cite{g3} for
the  extremal Kerr-Newman and Kerr-Newman-AdS-dS black holes. The
Casimir element of $so(2,1)$ is given in terms of $L$ and $q$ \be
{\cal I}=HK-D^2=L(\t,p_\t,p_\varphi)-q(p_\varphi)^2. \ee
 Note that
the angular sector of the system under consideration defined by the
 Hamiltonian  system
 \be\left({\cal I}(\theta, p_\theta, p_\varphi),\;\omega_0=dp_\theta\wedge
d\theta+dp_\varphi\wedge d\varphi\right),\ee is an integrable system. Thus,
one can introduce the action-angle variables $(I_a,\Phi^a)$ :
 \be \left({\cal I}=L(I_1,I_2)-q(I_2)^2,\;
\omega_0=dI_a\wedge d\Phi^a\right),\qquad \Phi^a\in[0,2\pi), \quad a=1,2.
\label{sph} \ee
In these terms the conformal generators acquire the
form \be \label{haa} H={r}\left(\sqrt{{(r p_r)}^2 +L(I_{1},I_2) }
-q(I_2)\right),\quad K=\frac{1}{r} \left(\sqrt{ {(r p_r)}^2
+L(I_{1},I_2) } +q(I_2)\right), \quad D=r p_r. \ee

At this point, the Hamiltonian can be put in the conventional
conformal mechanics form by introducing the new radial coordinates as in \cite{hkln}
\be\label{xp}
R=\sqrt{2 K}, \qquad P_R=-\frac{2 D}{\sqrt{2 K}}, \qquad \{R,P_R\}=1 \quad \Rightarrow \quad H=\frac 12 P^{2}_R+\frac{2 {\cal I}}{R^2},
\ee
with ${\cal I}$ from (\ref{sph}). However, with respect to the Poisson bracket the new radial variables $(R, P_R)$ do not commute
with $\Phi^a$.
In order to split them,
we replace the angle variables by the following ones
\be
{\widetilde\Phi^a}=\Phi^a+\frac12\int d({RP_R})\frac{\partial\log\left(\sqrt{(RP_R)^2/4 +L(I_1,I_2) } +q(I_2)\right)
}{\partial I_a}. \label{tphi}\ee
In these terms
the symplectic structure takes the canonical form
\be
\Omega=dP_R\wedge dR+dI_a\wedge d{\widetilde\Phi}^a, \ee
i.e. $(R,P_R)$ and $(\tilde \Phi^a,I_a)$ constitute canonical pairs.

Thus, we defined a canonical transformation that transforms the model  of
a massive relativistic particle moving near the horizon
of an extremal four-dimensional black hole to the conventional conformal mechanics.
However, this transformation assumes the action-angle variables of the angular sector of the system.

Let us remind, that for the construction of action angle variables for the system $(\omega_0, {\cal I})$,
\be
{\cal I}=L(\theta,p_\theta, p_\varphi )-q(p_\varphi)^2,\qquad \omega_0=dp_\theta\wedge d\theta+ dp_\varphi\wedge d\varphi,
\ee
we should introduce the generating function \cite{arnold}
\be
S({\cal I}, p_\varphi, \theta, \varphi )
= p_\varphi \varphi
+\int_{\scriptsize{\begin{array}{c}
{\cal I}={\rm const}\\
 p_\varphi={\rm const}
 \end{array}}} p_\theta({\cal I}, p_\varphi, \theta) d\theta = p_\varphi \varphi + S_0({\cal I}, p_\varphi, \theta).
\label{s0}\ee
Then, we define, by its use, the action-angle variables
\be
I_{1}({\cal I}, p_\varphi)=\frac{1}{2\pi}\oint p_\theta({\cal I}, p_\varphi , \theta) d\theta,\quad I_2=p_\varphi,\qquad
\Phi_{1,2}=\frac{\partial{S({\cal I}(I_1,I_2), I_2, \theta, \varphi )}}{\partial I_{1,2}},
\ee
where ${\cal I}(I_1,I_2)$ is obtained from the first expression above.

\section{Reissner-Nordstr\"om black hole}
In this section we construct the action-angle variables for the angular part of the conformal mechanics describing the motion of charged particle near
horizon of extreme Reissner-Nordstr\"om black hole (which defines the electrically and magnetically charged static black hole configuration)

The conformal generators which characterize the charged probe particle  read
\cite{galaj}
\bea\label{h}
&&
H=\frac{r}{M^2}\left(\sqrt{(mM)^2+{(rp_r)}^2 +p_\t^2+\sin^{-2}\t
(p_\vf + ep\cos\theta)^2} +e q\right), \qquad D=r p_r,
\nonumber\\[2pt]
&&
K=\frac{M^2}{r} \left(\sqrt{{(mM)}^2+{(rp_r)}^2
+p_\t^2+\sin^{-2}\t (p_\vf+ep\cos\theta )^2} -e q\right),
\eea
where $m$ and $e$ are the mass and the electric charge of a particle,
while $M$ , $q$ and $p$ are, respectively,  the mass,  the electric
charge, and the magnetic charge of the black hole.
From these expressions we immediately get
 angular part of our system
\be
{\cal
I}=p^2_\theta+\frac{(p_\varphi+s\cos\theta)^2}{\sin^2\theta}+(mM)^2-(eq)^2,\quad
\omega = dp_\theta\wedge d\theta+dp_\varphi\wedge d\varphi.
\label{br}\ee
It is precisely the spherical Landau problem (Hamiltonian system, describing the motion of the particle on the sphere
in the presence of constant magnetic field generated by Dirac monopole), shifted on the constant
${\cal I}_0 =(mM)^2-(eq)^2$.
Here and through this section we will use the notation
\be
s=ep\;,
\ee
which is precisely the  Dirac's ``monopole number".

For the construction of action-angle variables of the obtained system,
let us introduce the generating function (\ref{s0}), where the second term looks as follows
\be  S_0({\cal I}, p_\varphi , \theta)= \int_{{\cal I}={\rm
const}} d\theta\sqrt{{\cal I}-
(mM)^2+(eq)^2-\frac{(p_\vf + s\cos\theta)^{2}}{\sin^2\theta}}=\nonumber\ee
 \be
=\left\{
\begin{array}{cc}
2|s|\arcsin\frac{|s|}{\sqrt{{\widetilde {\cal I}}}} \cot\frac\theta2-2\sqrt{{\widetilde {\cal I}}+s^2}\arctan \frac{\sqrt{{\widetilde {\cal I}}+s^2}\cot\frac\theta2}{\sqrt{{\widetilde {\cal I}}-s^2\cot^2\frac\theta2}}
&{\rm for}\; p_\varphi=s\\
-2|s|\arcsin\frac{|s|}{\sqrt{{\widetilde {\cal I}}}} \tan\frac\theta2+2\sqrt{{\widetilde {\cal I}}+s^2}\arctan \frac{\sqrt{{\widetilde {\cal I}}+s^2}\tan\frac\theta2}{\sqrt{{\widetilde {\cal I}}-s^2\tan^2\frac\theta2}}
&{\rm for}\; p_\varphi=-s\\

\sqrt{{\widetilde {\cal I}}+s^2} \left[
2\arctan t-\sum_{\pm}
\sqrt{((1\pm b)^2- a^2)}\arctan{\frac{(1\pm b)t\pm a}{\sqrt{(1\pm b)^2-a^2}}}\right],&{\rm for}\; |p_\varphi|\neq |s|
\end{array} \right. .
\ee
Here we introduce the notation
\be
{\widetilde {\cal I}}={\cal I}-(m M)^2+(e q)^2,\quad
a^2\equiv\frac{{\widetilde{\cal I}}^2 -(p^2_\varphi-s^2){\widetilde{\cal I}}}{({\widetilde{\cal I}}+s^2)^2},
 \quad b\equiv-\frac{s p_\varphi}{{\widetilde{\cal I}}+s^2},
\quad
t=\frac{a-\sqrt{a^2-(\cos\theta -b)^2}}{\cos\theta -b}.
\ee
{\sl Hence, the equation $p_\varphi=\pm s$ defines critical points, where the system  changes its behaviour.}

For the non-critical values $p_\varphi \neq \pm s$ we get, by the use of standard methods \cite{arnold}, the
following expressions for the action-angle variables,
\be
I_1=\sqrt{{\widetilde{\cal I}}+s^2} -\frac {| s+p_\varphi| +| s-p_\varphi |}{2},\qquad
\Phi_1=-\arcsin
\frac{({\widetilde{\cal I}}+s^2)\cos\theta+s p_\varphi}{{\widetilde{\cal I}}^2 -(p^2_\varphi-s^2){\widetilde{\cal I}}}
\label{actions1}
\ee
\bea &&
I_2=p_\varphi,\qquad \Phi_2=\varphi+\gamma_1\Phi_1
+\gamma_2 \arctan\left(\frac{a-(1-b)t}{\sqrt{(1-b)^2-a^2}}\right)
+\gamma_3 \arctan\left(\frac{a+(1+b)t}{\sqrt{(1+b)^2-a^2}}\right)
\eea

where
\be
(\gamma_1,\gamma_2,\gamma_3)=\left\{
\begin{array}{lrrr}
sgn(I_2)(1,&-1,&1)& {\rm for }\;|I_2|>|s|\\
sgn(s)(0,&-1,&-1)& {\rm for }\;|I_2|<|s|
\end{array}
\right.
\ee
Respectively, the Hamiltonian reads
$$
{\cal I}=\left(I_1+\frac {| s+I_2| +| s-I_2 |}{2}\right)^2
+(mM)^2-(eq)^2-s^2=
$$
\be
=\left\{\begin{array}{cc}
(I_1+|I_2|)^2+(mM)^2-(eq)^2-s^2 &{\rm for }\; |I_2|>|s|\\
(I_1+|s|)^2+(mM)^2-(eq)^2-s^2 &{\rm for }\; |I_2|<|s|\\
\end{array}\right.
\ee

The effective frequencies $\Omega_{1,2}=\partial {\cal I}/\partial I_{1,2}$ looks as follows
\be
{\Omega_1}=\left\{\begin{array}{cc}
2(I_1+|I_2|) &{\rm for }\; |I_2|>|s|\\
2(I_1+|s|) &{\rm for }\; |I_2|<|s|
\end{array}\right. ,
\qquad
{\Omega_2}=\left\{\begin{array}{cc}
2 (I_1+|I_2|) {\rm sgn} I_2&{\rm for }\; |I_2|>|s|\\
0 &{\rm for }\; |I_2|<|s|\\
\end{array}\right. .
\ee
It is seen, that  in subcritical  regime, $|I_2|< s $, the frequency $\Omega_2$ becomes zero, while
frequency $\Omega_1$  depends on $I_1$ only. In overcritical regime, when $|I_2|>s $, the frequencies $\Omega_1$ and $\Omega_2$
coincides modulo to sign:the frequency $\Omega_2$ is positive
for positive values of $I_2$ (which is precisely  angular momentum $p_\varphi $), and vice versa.
This is essentially different from the periodic motion in the spherical part of the ``Kerr particle" observed in \cite{bny}, where the critical point separated two phases, both of which corresponded to the two-dimensional motion, but with opposite sign of $\Omega_2$.

So, in both regimes the trajectories are closed, and the motion is effectively one-dimensional one.
It reflects the existence of the additional constant of motion in the system  (\ref{br}),
reflecting the  $so(3)$ invariance of the spherical Landau problem.
  In other words, it is superintegrable one.
 In action-angle variables the additional constant of motion reads
$I_{add}=\sin(\Phi_1-\Phi_2)$ (cf.\cite{gonera,hlnsy}).
\\
Now, let us write down the expressions for action-angle variables at the critical point $p_\varphi=\pm s$,
\be
I_1=2({\sqrt{\widetilde{\cal I}+s^2}}-|s|)
\;,\quad
\Phi_1=\left\{
\begin{array}{cc}
-2\arctan \frac{\sqrt{\widetilde{\cal I}+s^2}\cot\frac\theta2}{\sqrt{{\widetilde{\cal I}}-s^2\cot^2\frac\theta2}}
&{\rm for}\; p_\varphi=s\\
 2\arctan \frac{\sqrt{\widetilde{\cal I}+s^2}\tan\frac\theta2}{\sqrt{{\widetilde{\cal I}}-s^2\tan^2\frac\theta2}}
&{\rm for}\; p_\varphi=-s
\end{array} \right.\ee
Respectively, the Hamiltonian reads
\be
{\cal I}=(\frac{I_1}{2}+|s|)^2+(m M)^2-(e q)^2-s^2
\ee
Notice, that the obtained action variable is not the corresponding limit of (\ref{actions1}).

\section{Cl\'ement-Gal'tsov black hole}
Now, let us  consider the motion of a particle near the horizon of extremal rotating
Cl\'ement-Gal'tsov(dilaton--axion) black hole \cite{cg1}.
The conformal generators of this particle system (with mass $m$ and ``effective monopole number" $s$,which was refereed in \cite{cg,cg1} as  ``effective electric charge" $e$) read
\cite{cg}
\bea\label{h1}
&&
H=r \left(\sqrt{m^2+{(rp_r)}^2 +p_\t^2+\sin^{-2}\t
{[p_\vf-s\cos\t]}^2   } -p_\vf \right), \qquad D=r p_r,
\nonumber\\[2pt]
&&
K=\frac{1}{r} \left(\sqrt{m^2+{(rp_r)}^2 +p_\t^2+\sin^{-2}\t
{[p_\vf-s \cos\t]}^2   } +p_\vf \right).
\eea
The Casimir of  conformal algebra is given by the expression
\be
{\cal I}=p^2_\theta +\frac{(p_{\varphi}\cos\theta -s)^2}{\sin^2\theta}+m^2.
\label{calIscal}\ee
The second term in  generating function for the action-angle variables (\ref{s0})
 can be explicitly integrated in elementary functions
(its explicit expression could be found, e.g. in Appendix in \cite{aa})
 \be   S_0=\int_{{\cal I}={\rm
const}} d\theta\sqrt{{ \cal I}-
m^2-\frac{(p_{\varphi}\cos\theta -s)^2}{\sin^2\theta}}=
\nonumber\ee
\be
=\left\{
\begin{array}{cc}
-2|s|\arcsin\frac{|s| \tan\frac\theta2}{\sqrt{{\cal I}-m^2}}+2\sqrt{{\cal I}-m^2+s^2}\arctan \frac{\sqrt{{\cal I}-m^2+s^2}
\tan\frac\theta2}{\sqrt{{\cal I}-m^2-s^2\tan^2\frac\theta2}}
&{\rm for}\; p_\varphi=s\\
2|s|\arcsin\frac{|s| \cot\frac\theta2}{\sqrt{{\cal I}-m^2}}-2\sqrt{{\cal I}-m^2+s^2}\arctan \frac{\sqrt{{\cal I}-m^2+s^2}
\cot\frac\theta2}{\sqrt{{\cal I}-m^2-s^2\cot^2\frac\theta2}}
&{\rm for}\; p_\varphi=-s\\
\sqrt{{\cal I}-m^2+p^2_\varphi} \left[
2\arctan t-\sum_{\pm}
\sqrt{((1\pm b)^2- a^2)}\arctan{\frac{(1\pm b)t \pm a}{\sqrt{(1\pm b)^2-a^2}}}\right],&{\rm for}\; |p_\varphi|\neq |s|
\end{array} \right.\ee
where we introduce the notation
\be
a^2\equiv\frac{({\cal I}-m^2)^2 +({\cal I}-m^2)(p^2_\varphi-s^2)}{({\cal I}-m^2+p^2_\varphi )^2},
\qquad b\equiv\frac{s p_\varphi}{{\cal I}-m^2+p^2_\varphi},
\qquad
t=\frac{a-\sqrt{a^2-(\cos\theta -b)^2}}{\cos\theta -b}.
\ee
{\sl Hence, the equation $p_\varphi=\pm s $ defines critical points, where the system  changes its behaviour.}

For the non-critical values $p_\varphi \neq \pm s$ we get, by the use of standard methods \cite{arnold}, the
following expressions for the action-angle  variables,
\be
I_1=\sqrt{{\cal I}-m^2+p^2_\varphi} -\frac {| s+p_\varphi| +| s-p_\varphi |}{2},\qquad\Phi_1=-\arcsin
\frac{({\cal I}-m^2+p^2_\varphi)\cos\theta-s p_\varphi}{\sqrt{({\cal I}-m^2)^2 +({\cal I}-m^2)(p^2_\varphi-s^2)}}
\label{actions2}\ee
\bea &&I_2=p_\varphi,\qquad
\Phi_2=\varphi+\gamma_1\Phi_1
+\gamma_2 \arctan\left(\frac{a-(1-b)t}{\sqrt{(1-b)^2-a^2}}\right)
+\gamma_3 \arctan\left(\frac{a+(1+b)t}{\sqrt{(1+b)^2-a^2}}\right)
\eea
where
\be
(\gamma_1,\gamma_2,\gamma_3)=\left\{
\begin{array}{lrrr}
sgn(I_2)(1,&-1,&1)& {\rm for }\;|I_2|>|s|\\
sgn(s)(0,&1,&1)& {\rm for }\;|I_2|<|s|
\end{array}
\right.
\ee
Respectively, the Hamiltonian reads
\be
{\cal I}=\left( I_1+\frac{|s+I_2| +|I_2 -s|}{2} \right)^2-I^2_2+m^2 =\left\{\begin{array}{cc}
(I_1+|I_2|)^2-I^2_2+m^2 &{\rm for }\; |I_2|>|s|\\
(I_1+|s|)^2-I^2_2+m^2 &{\rm for }\; |I_2|<|s|\\
\end{array}\right. .
\ee

In the critical points $p_\varphi=\pm s$, the action-angle  variables read
\be
I_1=2(\sqrt{{\cal I}-m^2+s^2}-|s|),\qquad\Phi_1=\left\{
\begin{array}{cc}
 2\arctan \frac{\sqrt{{\cal I}-m^2+s^2}\tan\frac\theta2}{\sqrt{{\cal I}-m^2-s^2\tan^2\frac\theta2}}
&{\rm for}\; p_\varphi=s\\
-2\arctan \frac{\sqrt{{\cal I}-m^2+s^2}\cot\frac\theta2}{\sqrt{{\cal I}-m^2-s^2\cot^2\frac\theta2}}
&{\rm for}\; p_\varphi=-s
\end{array} \right.
\ee
Inverting the first expression, we shall get the expression for Hamiltonian
\be
{\cal I}=(\frac{I_1}{2}+|s|)^2+m^2-s^2
\ee
Let us notice, that the action variable at the critical point is different from  the corresponding limit of (\ref{actions2}).

To clarify the meaning of critical point let us calculate the effective frequencies  of the system,
 $\Omega_{1,2}=\partial {\cal I}/\partial I_{1,2}$,
\be
{\Omega_1}=\left\{\begin{array}{cc}
2(I_1+|I_2|) &{\rm for }\; |I_2|>|s|\\
2(I_1+|s|) &{\rm for }\; |I_2|<|s|\\
\end{array}\right. ,
\qquad
{\Omega_2}=\left\{\begin{array}{cc}
2I_1sgnI_2&{\rm for }\; |I_2|>|s|\\
-2 I_2 &{\rm for }\; |I_2|<|s|\\
\end{array}\right. .
\ee
It is seen from this expressions, that in contrast
with previous case, the system does not possess the hidden symmetries.
The motion is nondegenerated in noncritical regimes.
So, in contrast with Reissner-Nordstr\"om case, the system is essentially
two-dimensional one, and its trajectories are unclosed.
The frequencies $\Omega_1$ are the same in the both cases, while  $\Omega_2$ are essentially different.
Moreover,  frequency $\Omega_2$ behaves in essentially different ways in subcritical and overcritical regimes.
 In the first case it is proportional to $I_1$, and in the second case to $I_2$.

{\large Acknowledgements.}
I am indebted to Anton Galajinsky and Armen Nersessian for their kind suggestion to consider  this problem.
Special thanks to Armen Nersessian for advice and for many useful discussions. This work was
supported  by the Volkswagen Foundation grant I/84 496,  by the
Armenian State Committee of Science  grant 11-1c258 and by the ANSEF grant 2774.

\end{document}